\documentstyle[preprint,psfig,aps]{revtex}
\tightenlines
\begin{document}
%\draft
\preprint{LBNL-42280}
\title{Dissociation rates of $J/\psi$'s with comoving mesons\\
--- thermal vs. nonequilibrium scenario}
\author{C.~Spieles$^1$\footnote{Supported by the Alexander v.~Humboldt Foundation}\footnote{Email: cspieles@lbl.gov },
R.~Vogt$^{1,2}$, L.~Gerland$^3$, S.A.~Bass$^4{}^*$, M.~Bleicher$^3$, H.~St\"ocker$^3$, W.~Greiner$^3$}

\address{$^1$~Nuclear Science Division,
Lawrence Berkeley National Laboratory,
Berkeley, CA 94720, USA}
\address{$^2$~Physics Department,
University of California at Davis, 
Davis, CA 95616, USA}
\address{$^3$~Institut f\"ur
Theoretische Physik,  J.~W.~Goethe-Universit\"at,
D-60054 Frankfurt a.M., Germany}
\address{$^4$~Department of Physics, Duke University,
Durham, N.C. 27708-0305, USA}
\maketitle

\begin{abstract}
We study $J/\psi$ dissociation processes in hadronic environments.
The validity of a thermal meson gas ansatz is tested by confronting it
with an alternative, nonequilibrium scenario.
Heavy ion collisions are simulated in the framework of the microscopic 
transport model UrQMD, taking into account the production of
charmonium states through hard parton-parton interactions and subsequent
rescattering with hadrons.
The thermal gas and microscopic transport scenarios are shown to be very
dissimilar. Estimates of $J/\psi$ survival probabilities based on thermal
models of comover interactions in heavy ion
collisions are therefore not reliable.
\end{abstract}

\newpage

\section{Introduction}
The suppression of heavy quarkonia has long been proposed as a unique
signature for the transient creation of a quark-gluon plasma in heavy ion
collisions \cite{matsui}. Preliminary reports on $J/\psi$ and $\psi'$
production cross sections in Pb+Pb reactions at SPS energies
\cite{abreu,romana} in fact seem to indicate an anomalous charmonium
suppression compared to $pA$ and lighter $AB$ interactions. 
It has been demonstrated that the framework of Glauber theory 
with some comover interactions \cite{kharzeev,vogt98} fails to describe 
all the data while assuming the onset of deconfinement at a certain interaction 
density leads to a better agreement with experiment. 
Starting from a similar ansatz, however, it was
claimed that hadronic mechanisms alone could account for the observed
suppression probabilities \cite{armesto}. 
While in these studies universal comover absorption cross sections are 
parameters, there have been attempts to predict the actual comover
cross sections for certain reaction channels and to estimate the resulting
$J/\psi$ survival probabilities \cite{martins,mueller}. From these studies
one learns that the dissociation cross sections may depend strongly on
the scattering energy as well as the
meson species interacting with the charmonium state. 
Unfortunately only {\em thermal} absorption rates have been calculated 
on the basis of the predicted cross sections. 

It is the goal of the present work to investigate, on the basis of a
hadronic transport model, if the assumption of 
a thermal hadronic environment can be justified. 
This is analogous to studies of the nonequilibrium processes
contributing to strangeness production \cite{mat89} versus thermal model
calculations \cite{koch}.
Scattering rates 
and collision energies in the microscopic simulation are compared to the
ideal hadron gas scenario. We also address the question of whether only the
consideration of light mesons ($\pi$ and $\rho$) suffices to account 
for the major part of all possible comover absorption processes.

\section{The model}
The production and subsequent absorption of charmonium states in relativistic heavy
ion collisions are the result of an intricate interplay between 
hard and soft processes.
Since the interactions are nonperturbative the calculation of their 
dynamics from first principles is prohibitive.
Thus phenomenological models which account for both 
hadronic and partonic aspects are needed. 
We simulate charmonium production and absorption microscopically.

The charmonium states are produced exclusively in hard parton-parton
interactions for 
which a simple Glauber-type picture of nucleus-nucleus reactions applies.
This is supported by the fact that the experimentally measured
dimuon continuum in the corresponding
mass range can be well described by perturbative calculations of the
Drell-Yan process\footnote{The production of $c\bar c$ pairs which eventually
form the quarkonium state is, however, dominated by gluon-gluon
scatterings.}. In particular, a linear scaling of the Drell-Yan cross
section with atomic mass $A$ is 
observed in $pA$ interactions~\cite{Al90}. 
Initial state interactions are assumed to be negligible. 
We apply perturbative QCD to nucleus-nucleus collisions
without nuclear modifications of the parton distribution functions. (See
Refs.~\cite{Eme98,Ham98} for the effects of such modifications.)

Within perturbative QCD, the hard production process is 
completely decoupled from other possible interactions such as the 
creation and reinteraction of soft quarks and gluons and their rearrangement 
into color singlet states. In particular, the hard production process
is not linked to soft hadron production, baryon stopping and hadronic 
rescattering.

The dissociation of charmonium states is assumed to be due to 
hadronic interactions. The space-time evolution of
mesons and baryons, governed by soft physics, can be treated by
transport theory on the hadronic level. 
The kinetics of each interaction is determined by appropriate hadronic 
scattering cross sections. In particular, charmonium dissociation can be 
studied by introducing cross sections for their absorption through 
interactions with baryons and mesons. In this work, we consider dissociation
by comoving mesons only\footnote{The calculation of {\em nuclear}
absorption within our model framework shows interesting deviations
from the standard Glauber-like prescription. Due to nuclear stopping in the
UrQMD simulation, incident nucleons are slightly retarded which leads to an
increase of the effective path length of $J/\psi$'s in nuclear matter.
This will be discussed in \cite{prep}.}.

Technically, we generate a space-time distribution of charmonium production 
points for any impact parameter $b$ by
microscopically simulating Glauber-type nucleus-nucleus collisions
assuming that
the nuclei, {\em i.e.} sets of appropriately initialized nucleons, 
pass through each other on straight line trajectories without energy loss. 
At the space-time point of each nucleon-nucleon encounter, a charmonium state
has a fixed production cross section.

To calculate the rescattering of charmonium states we generate a full
hadronic cascade simulation with the Ultrarelativistic Quantum Molecular
Dynamics, UrQMD, model \cite{bigpaper} which takes stopping and
particle production into account. 
The charmonia are inserted into
the evolving hadronic environment at the appropriate space-time points
according to the Glauber simulation of the nucleus-nucleus collision.
The momenta of the charmonium states are assigned from
the parametrization \protect\cite{ramona}: 
\[
E\frac{d\sigma}{dMdp^3}\sim(1-x_F)^{3.55} \exp(-p_T\, 2.08\rm \,GeV^{-1})\;
.
\]
In principle, the total energy of the produced $c\bar c$ state
has to be subtracted from the overall reaction. However, it is
difficult to do so consistently since simulations of the
charmonium production and rescattering processes are formulated within
different conceptual frameworks. 
Therefore we allow for the small violation of energy-momentum conservation
imposed by this procedure.
In Pb+Pb collisions at 160~GeV/nucleon,
$\sqrt{s}=A\sqrt{s}_{NN}\approx 3600$~GeV and with a typical 
charmonium energy of $E_{c\bar c}\approx 4$~GeV,
$E_{c\bar c}/\sqrt{s}\approx 0.1$\% . 

In the simulations we have used fixed dissociation cross sections,
$\sigma_{J/\psi M}= 2 \sigma^{\rm tot}_{J/\psi N}/3$ according to the 
additive quark model, for any meson species $M$.
The value of the total $J/\psi N$ absorption cross section, taken from
Ref.~\cite{gerland}, is $\sigma^{\rm tot}_{J/\psi N}=3.62$~mb.
Since we do
not consider elastic interactions, any collision above the kinematic
threshold leads to the dissociation of
the $J/\psi$ so that we use the two terms synonymously.
For this study, we have neglected the energy dependence of the dissociation
cross sections which may be quite strong close to threshold. 
Also, we have not considered a possible (eigen)time
dependence of the actual dissociation cross sections which may evolve from
a small initial cross section at the production point
to its asymptotic value (see \cite{gerland}). We want to focus on the
question of whether or not the $J/\psi$-comover reactions in a heavy ion collision
can be reasonably approximated by a thermal hadron gas
interacting with the $J/\psi$. The qualitative answer to this 
question is rather insensitive to the actual parametrization of the cross
section.

Note that we do include the formation time of the comovers
(on average, $\tau_F\approx 1$~fm/$c$).
Particles produced in a string fragmentation process 
are not allowed to interact with each other or with a $J/\psi$
within their formation time\footnote{It is conceivable that a nonvanishing
parton content of the yet-to-be-formed hadrons leads to a finite dissociation
cross section before $\tau_F$. 
Dimuon production has been shown to be enhanced by partonic interactions of
preformed mesons in a similar context \cite{epj}.}.
%A reduced cross section applies for leading
%hadrons which contain constituent (anti-)quarks from one of the incident
%nucleons. 

We assume $E_{th}=2m_D=3730$~MeV is the kinematic threshold for
$J/\psi$-meson dissociation
processes. In the thermal model, a fixed $\rho$ mass of 770~MeV is used, therefore
the minimum $J/\psi\,\rho$ collision energy is $m_{J/\psi}+m_\rho=3870\,{\rm
MeV}>E_{th}$.
In the UrQMD model however, the finite width of the 
$\rho$ meson, $\Gamma=151$~MeV, is taken into account. 
The $\rho$ meson mass distribution is selected from the 
Breit-Wigner resonance formula.
Therefore, in the microscopic model some of the
$J/\psi\,\rho$ interactions are below threshold while no correpsonding
threshold exists for the thermal model.

The results in the equilibrium scenario are easily obtained.
Consider an ideal gas of $J/\psi$'s and one meson species $M$ ($\pi$ or
$\rho$). 
We calculate the rate $R$ of dissociation processes 
for an arbitrarily chosen $J/\psi$ in the rest frame of the $J/\psi$: 
\[
R=g_M \int {\rm d}^3 p_M f(p_M)j(p_M)\sigma_{J/\psi M} \, ,
\]
where the spin and isospin degeneracy, $g_M$, is 3 for $\pi$'s and 9 for 
$\rho$'s. The mesons are distributed according to the Bose-Einstein
distribution,
\[
f(p_M)=\frac{1}{(2\pi)^3}(\exp(E_M/T)-1)^{-1}
\]
where $E_M=\sqrt{m_M^2+p_M^2}$ is
the total energy, $p_M$ is the momentum of mesons in the rest 
frame of the $J/\psi$ and $j(p_M)=p_M/E_M$
is the meson flux. The dissociation cross section has a fixed value,
$\sigma_{J/\psi M}=2.41$~mb.
The above integral can be rewritten in terms of the center of mass 
collision energy, $E_{cm}$,  
\[
R= \int {\rm d}E_{cm} \frac{g_M}{2\pi^2} p_M^2 
\frac{E_{cm}}{m_{J/\psi}} (\exp(E_M/T)-1)^{-1} \sigma_{J/\psi M}
= \int {\rm d}E_{cm}  r(E_{cm},T) \, ,
\]
where $r(E_{cm},T)$ is the (unnormalized) collision spectrum for a given temperature
$T$. Consequently, the average collision energy is given by
\[
<E_{cm}(T)>=\frac{1}{R}\int {\rm d}E_{cm} E_{cm}\, r(E_{cm},T)
\, .
\]

\section{Results}

\begin{figure}[b]
\vspace*{\fill}
\centerline{\psfig{figure=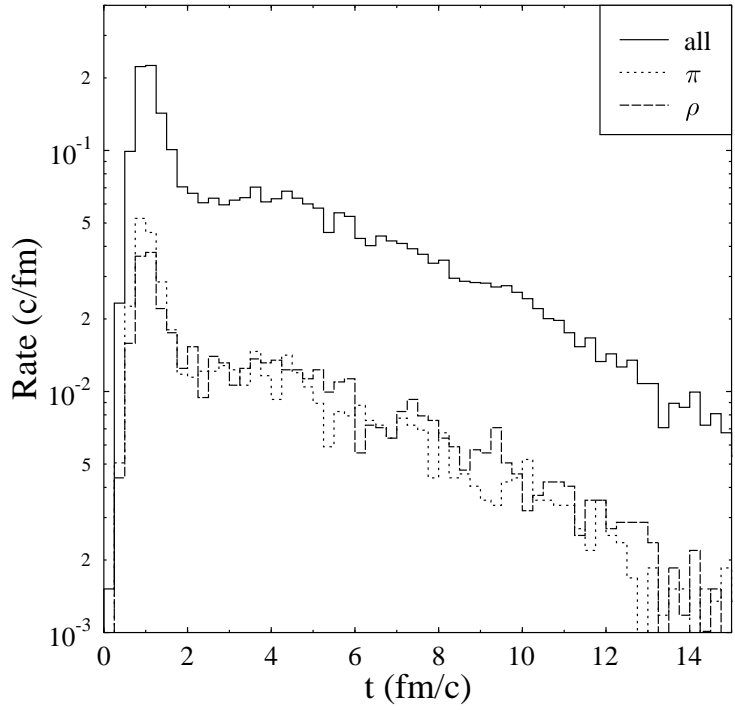,width=12cm}}
\caption{$J/\psi$ dissociation rates 
in central Pb(160~GeV)+Pb reactions according to the UrQMD calculation
as functions of time. Shown are the rates for all
$J/\psi$-meson interactions (full line), those for $J/\psi\,\pi$ (dotted line)
and those for $J/\psi\,\rho$ (dashed line). A fixed dissociation cross section
of $\sigma_{J/\psi M}=2.41$~mb is used for all mesons.
\label{rates}}
\vspace*{\fill}
\end{figure}

We first study very central Pb+Pb reactions ($b=0$) at $E_{\rm lab}=160$~GeV
with the microscopic nonequilibrium model.
Figure~\ref{rates} shows the $J/\psi$ dissociation rates as functions of time. 
Comover dissociation processes occur most frequently about 1~fm/$c$ after the
nuclear reaction begins. 
At that time, the total dissociation rate reaches a value of more than
0.1~$c$/fm.
However, it is dropping rapidly, an order of magnitude within the 
first 10~fm/$c$ of the reaction. The individual 
$J/\psi\,\pi$ and $J/\psi\,\rho$ interaction rates 
are plotted along with the total comover
absorption rates. It is clear that $\pi$'s and
$\rho$'s alone are not responsible for $J/\psi$ dissociation by comovers 
due to the large number of transient meson renonances 
present in the microscopic model. The string fragmentation
scheme used in UrQMD \cite{bigpaper} determines the relative 
abundancies of these states. This choice is far from unique 
since the model can be compared
to data only for the hadrons observable in the final state. 
In UrQMD, all states of eight different meson multiplets 
($J^{PC}=0^{-+},0^{++},1^{--},1^{+-},1^{++},2^{++}$ plus two excited
$1^{--}$ multiplets) may be populated in a string fragmentation process.
The probability to form a meson from one of these multiplets is chosen to be
proportional to the spin degeneracy and inversely proportional to the
average mass of the multiplet.

Table~\ref{table} shows the relative importance of the mesons which 
contribute most to $J/\psi$ dissociation. 
Scatterings of the $J/\psi$'s with $\pi$'s and $\rho$'s are
indeed the dominant dissociation processes. However, together they are
responsible for only 37\% of the total comover absorption. 
Twenty channels are left out of Table~\ref{table}, each of them 
contributing about 1\% or less. However, they account for 
more than 15\% of the total absorption. 

It is important to remember that the same dissociation cross section is used for
all $J/\psi$-meson interactions, most probably a crude approximation. 
For a better estimate of the $J/\psi$ dissociation rates, one would need to
calculate the cross sections in all possible channels. 
Of course, even if this whole set of cross sections was
completely determined it
is uncertain that the hadronic transport model describes all the details  of
rescattering correctly. 
In any case, it is questionable if a model of $J/\psi$-comover 
absorption which includes only the light mesons can be used for
reasonable quantitative predictions.

\begin{figure}[b]
\vspace*{\fill}
\centerline{\psfig{figure=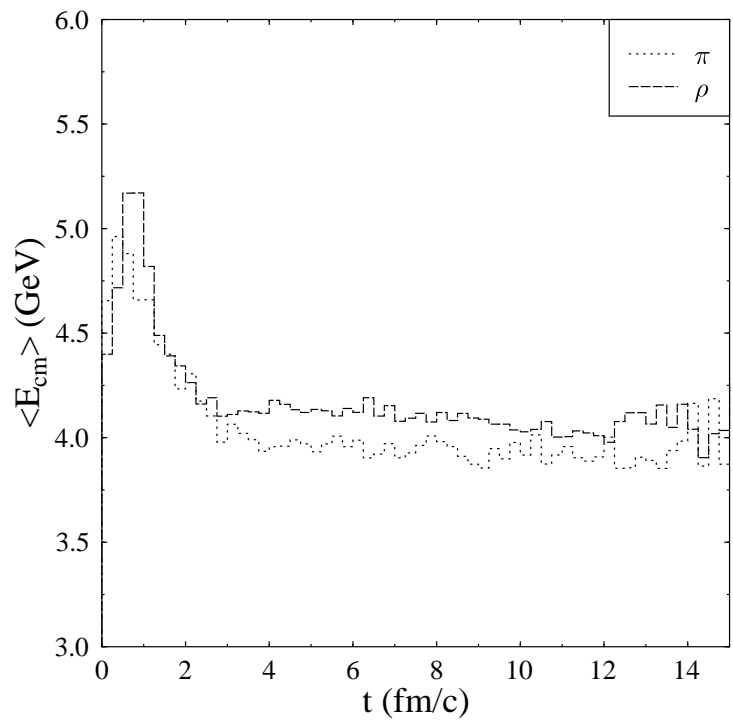,width=12cm}}
\caption{Average $J/\psi$-meson collision energies 
in central Pb(160~GeV)+Pb reactions as a function of time
for $J/\psi\,\pi$ (dotted line)
and $J/\psi\,\rho$ interactions (dashed line).
A fixed dissociation cross section of $\sigma_{J/\psi M}=2.41$~mb is used.
\label{ecmt}}
\vspace*{\fill}
\end{figure}

Figure~\ref{ecmt} shows the average $J/\psi$-meson collision energies
in central Pb(160~GeV)+Pb reactions as functions of time for
$J/\psi\,\pi$ and $J/\psi\,\rho$ interactions.
As one expects, the average collision energies are highest in the early
stage of the reaction where the rates are also at their maximum. 

\begin{figure}[b]
\vspace*{\fill}
\centerline{\psfig{figure=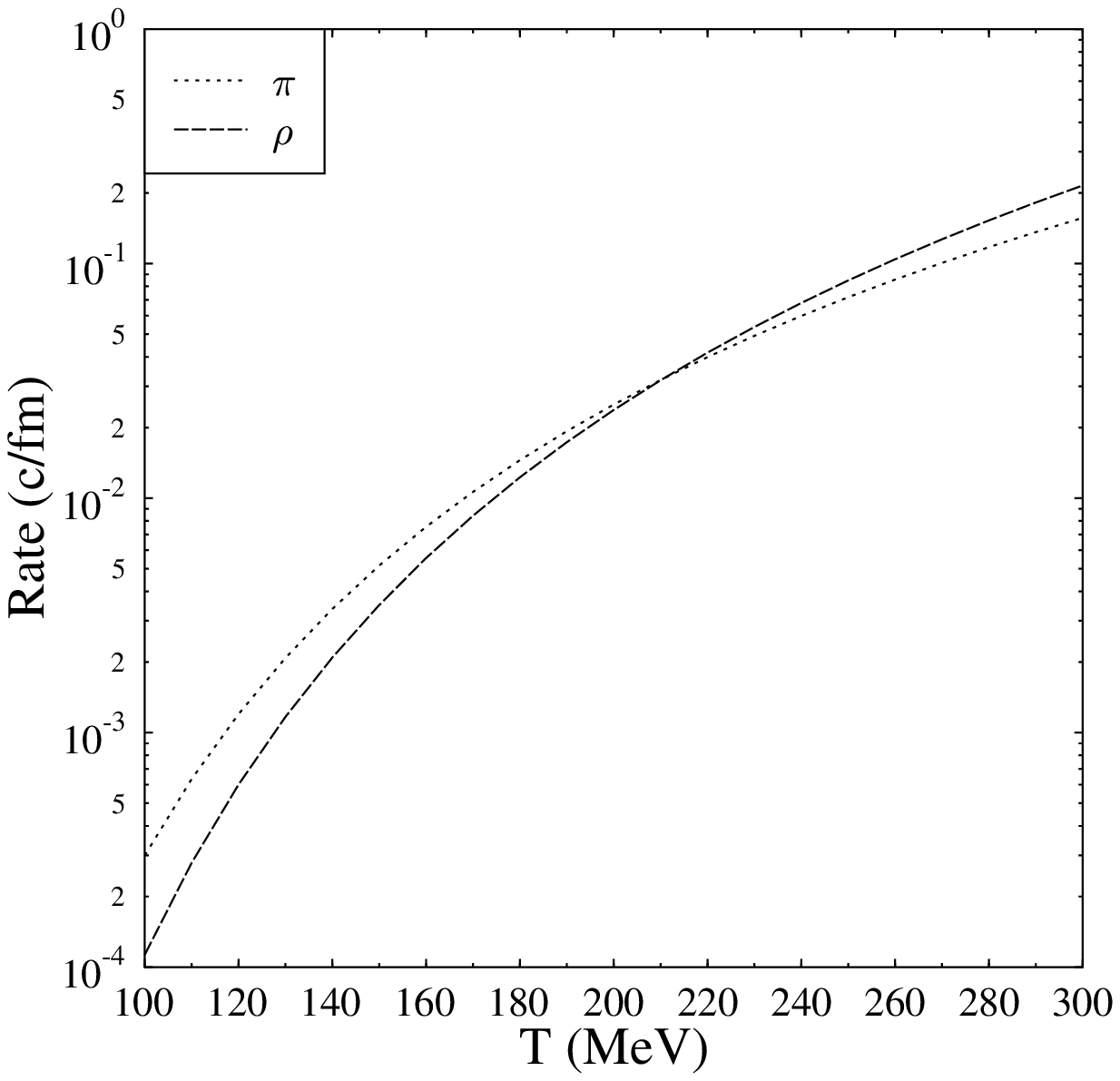,width=9cm}\hspace*{-1cm}
\psfig{figure=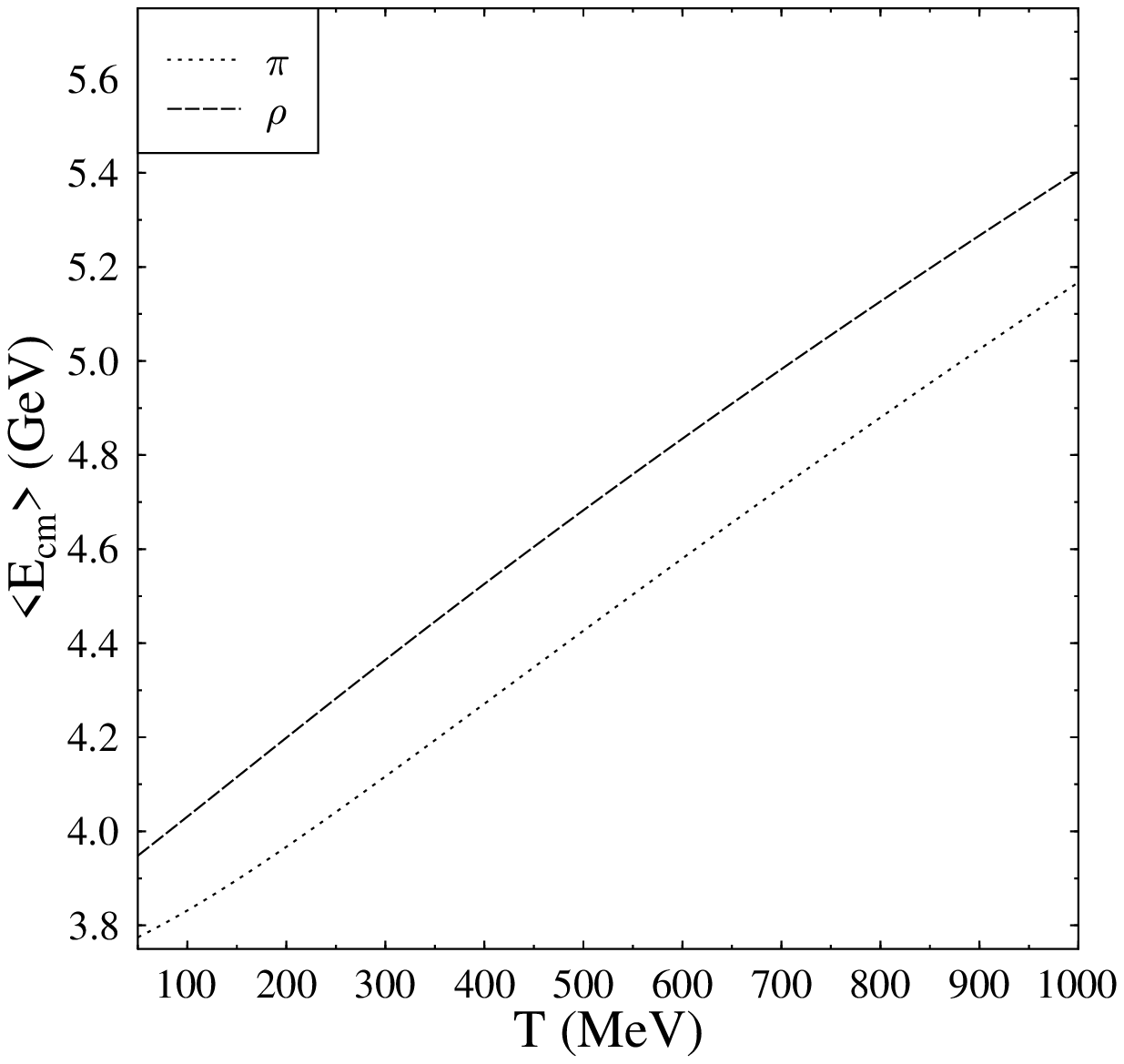,width=9cm}}
\caption{Left: $J/\psi$ dissociation rates in an ideal gas of pions and $\rho$'s
as a function of temperature. A fixed dissociation cross section
of $\sigma_{J/\psi M}=2.41$~mb is used.
Right: Average collision energies in the same system as a function of
temperature.
\label{therm_rates}}
\vspace*{\fill}
\end{figure}

We now compare the microscopic nonequilibrium scenario with a
thermal meson gas. Figure~\ref{therm_rates} shows the rates and the
averaged thermal dissociation energies in an ideal gas of
$\pi$'s and $\rho$'s as functions of temperature. The rates in the later
stage of the reaction, shown in Fig.~\ref{rates}, when $t>2$~fm/$c$, 
correspond
to a temperature of $T=140\pm 20$~MeV in the
equilibrated system. The temperatures one would deduce from the 
collision energies of Fig.~\ref{ecmt} roughly agree with these
values. Thus, the concept of a thermal hadron gas may be approximately valid
in the later stage of a nuclear reaction. The survival probability of
charmonium states, however, is to a large extent determined by the 
early reaction dynamics, $t\approx 1$~fm/$c$. Here, 
the rates roughly correspond to a thermal $\pi$ and $\rho$ 
gas with $T\approx 220$~MeV. In contrast, the average collision energies 
at this time are much higher than one could ever expect in a thermal hadron
gas. As can be read off from Fig.~\ref{ecmt} and Fig.~\ref{therm_rates},
the collision energies would correspond to temperatures greater than $800$~MeV. 
We can therefore conclude that a
major part of the $J/\psi$ dissociation processes cannot reasonably be
approximated by a thermal scenario.
Much higher collision energies are observed than can be expected
in a thermal system indicating that the underlying $\pi$ and $\rho$ momentum
distributions are out of equilibrium
and the probability of high relative velocities is enhanced. 

\begin{figure}[b]
\vspace*{\fill}
\centerline{\psfig{figure=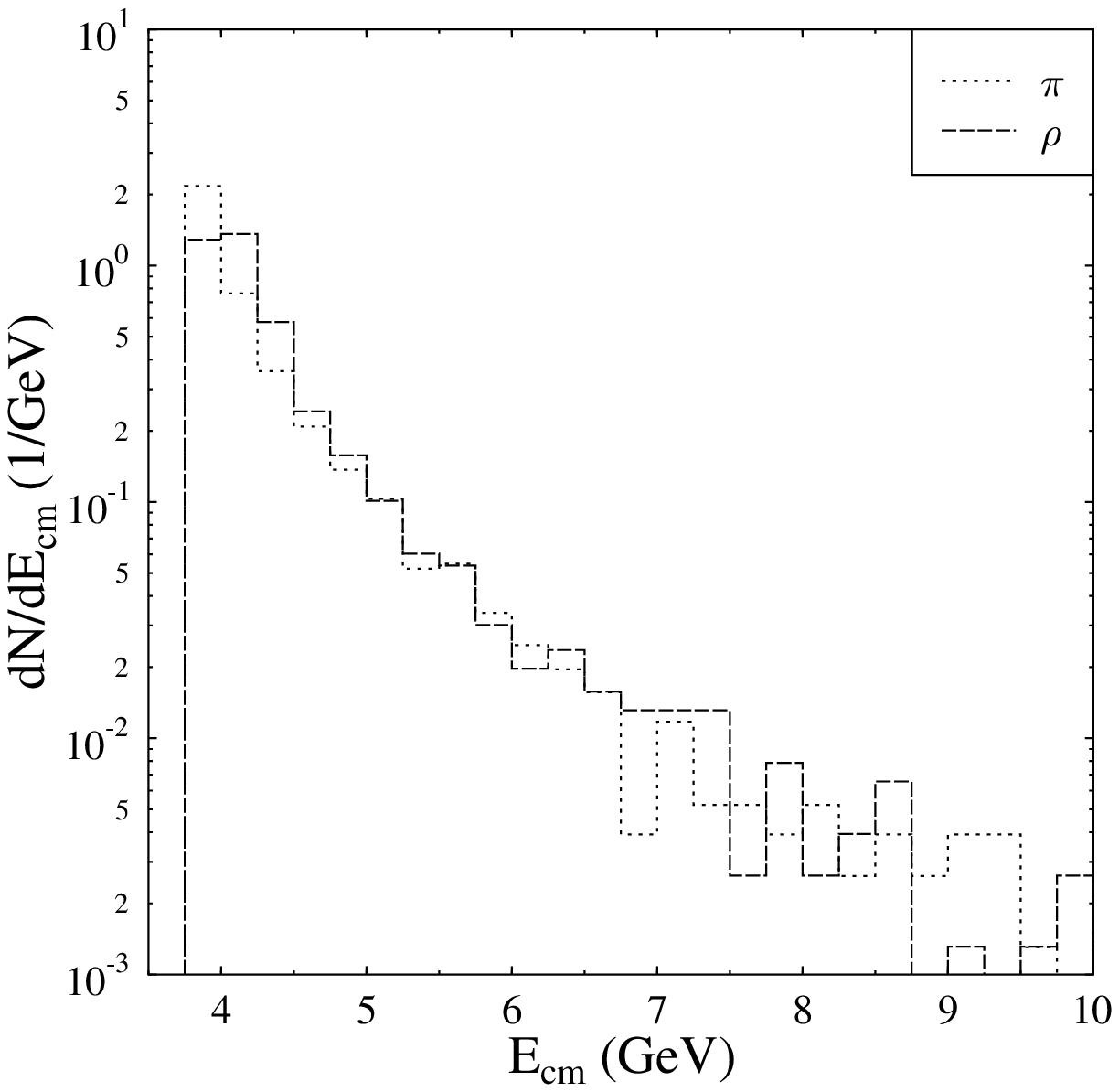,width=9cm}
\hspace*{-1cm}
\psfig{figure=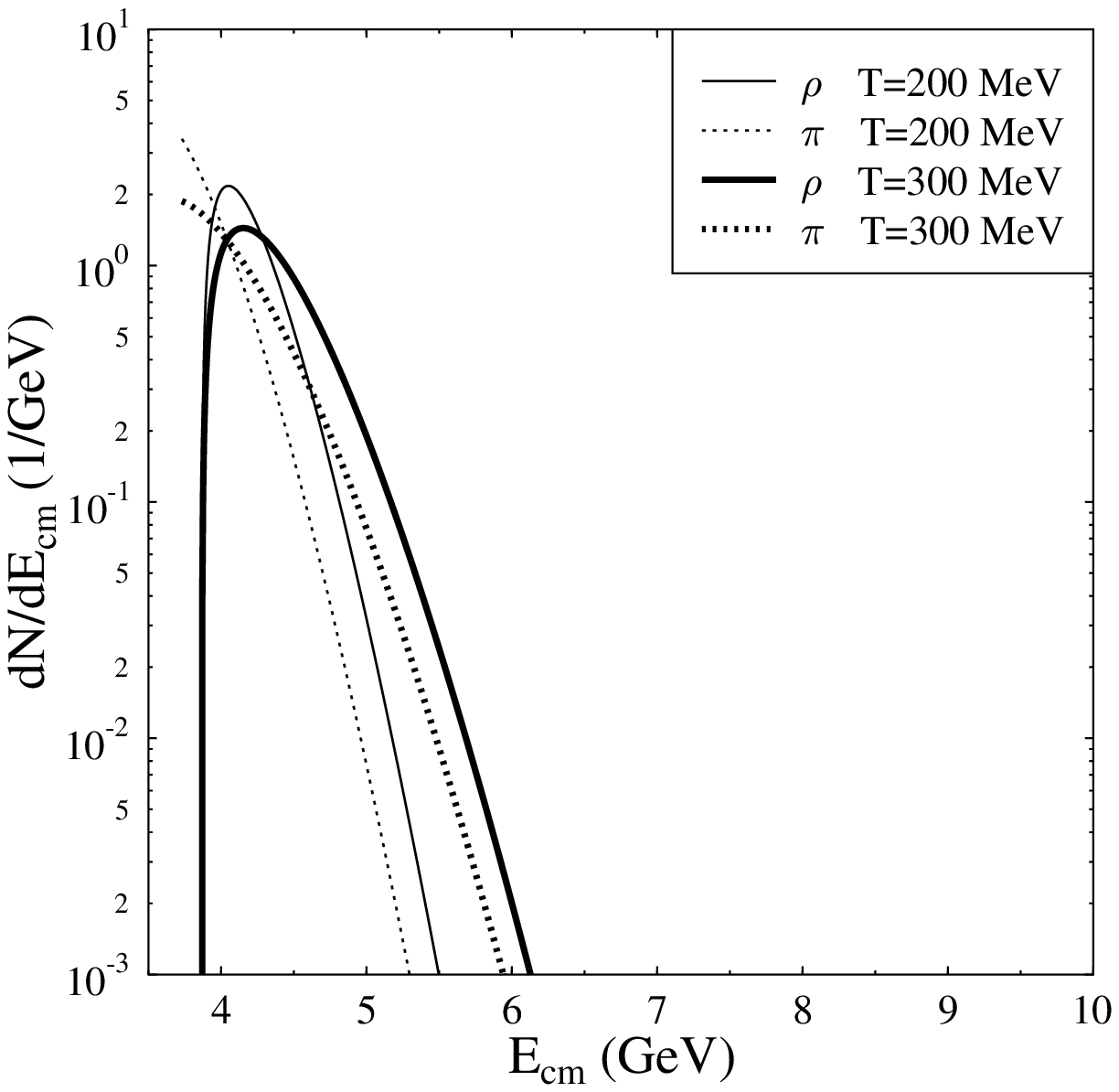,width=9cm}}
\caption{Left: $J/\psi$ collision spectrum
in central Pb(160~GeV)+Pb reactions according to the UrQMD calculation
(normalized to one collision).
Shown are the energy spectra for $J/\psi\,\pi$ interactions (dotted line)
and for $J/\psi\,\rho$ interactions (dashed line).
A fixed dissociation cross section of $\sigma_{J/\psi M}=2.41$~mb is used.
Right: $J/\psi$ collision spectra in an ideal gas of $\pi$'s and $\rho$'s
for $T=200$ and $300$~MeV, normalized to one collision. A fixed 
dissociation cross section of $\sigma_{J/\psi M}=2.41$~mb is used. Note that
the scales are identical in both plots.
\label{spec}}
\vspace*{\fill}
\end{figure}

The nonequilibrium character of $J/\psi$ rescattering is also
reflected by the time integrated collision energy spectrum. The prediction
of the nonequilibrium model is shown in Fig.~\ref{spec}, together with the
collision spectra of a thermal gas at $T=200$ and $300$~MeV.
Obviously, in the microscopic simulation collisions with high center of mass
energy are much more probable than in a thermal hadron gas at reasonable
temperatures.

As mentioned above, we have used fixed dissociation cross sections,
$\sigma_{J/\psi M}=2.41$~mb, in this
study. However, absorption by light mesons has recently been  
calculated in the framework of a meson exchange model \cite{mueller}
where the cross sections are strongly energy dependent. 
Thus, at a fixed comover density, the assumption of thermal velocity
distributions would considerably underestimate the
absorption rates. At $E_{cm}=5$~GeV, the energy dependent cross sections of Ref.~\cite{mueller}
are $\sigma_{J/\psi\,\pi}=2.3$~mb and
$\sigma_{J/\psi\,\rho}=1.6$~mb. As a result, using these cross sections 
\cite{mueller} in the nonequilibrium UrQMD simulation leads to slightly lower
maximum absorption rates at $t\approx 1$~fm/$c$ than do the fixed cross
sections. In the later stages of the reaction, the energy dependent 
absorption rates are 
more strongly suppressed since most of the collisions occur closer to
threshold, as shown in Fig.~\ref{ecmt}. Typically, for $t>3$~fm/$c$, the
collision energies are
$E_{cm}\approx 4$~GeV which implies 
$\sigma_{J/\psi\,\pi}\approx \sigma_{J/\psi\,\rho}\approx 0.3$~mb, considerably smaller than the cross sections
when $t\approx 1$~fm/$c$.

Certainly, different parametrizations of the energy and eigentime
dependencies for all possible comover dissociation cross sections would yield
different integrated survival probilities. In this work we explore 
only the influence of the comover dynamics on the absorption rates.
Comparison to experimental data with a refined prescription of charmonium
production and absorption in the microscopic framework will be presented elsewhere
\cite{prep}.

%The difference between the fixed dissociation cross section 
%$\sigma_{J/\psi M}=2.41$~mb ('AQM') and the energy dependent cross sections
%from \cite{mueller} are rather small at the early stage of the heavy 
%ion reaction where the average collision energies are large.
%At later stages, however, most of the $J/\psi$ meson collisions occur 
%at energies close to threshold which leads to a strong suppression in 
%the case of energy dependent cross sections.

\section{Conclusion}

We have presented a microscopic model of $J/\psi$ production and absorption in
relativistic heavy ion collisions. Averaged dissociation rates and collision
energies have been analyzed as functions of time. In this model, 
scattering of $J/\psi$'s
with $\pi$'s and $\rho$'s account for only about 37\% of the total
absorption by comovers. The $J/\psi$ absorption by comovers is 
strongly time dependent with dissociation
rates peaking at $t\approx 1$~fm/$c$. The collision energies at that time are
much higher than can be expected in a thermal scenario. 

An equilibrated gas of light mesons seems to be too simple an 
approximation to the actual
hadronic environment in which $J/\psi$'s interact.
In order to reliably calculate the survival probabilities of charmonium
states in heavy ion collisions, one must employ more realistic simulations 
of the meson dynamics. 
Many different reaction channels
contribute to the total absorption. The corresponding dissociation cross
sections and their energy dependencies need to be explored 
further with the help of phenomenological models.

\begin{table}
\caption{Contributions of different meson species to the total 
$J/\psi$ comover absorption according to the UrQMD simulation.
Only the most dominant channels are shown.}\label{table}
\begin{tabular}{|c|c|}  
\hline
Species &  Percentage of comover absorption \\ \hline
$\pi$  &  $18.5 \pm 0.3$ \\
$\rho$ &  $18.4 \pm 0.3$  \\
$K$     &  $10.7 \pm 0.3$  \\
$K^*(892)$ &  $7.4 \pm 0.2$ \\ 
$\eta$ &  $6.4 \pm 0.2$  \\
$\omega$ &  $6.4 \pm 0.2$  \\
$a_2(1320)$ &  $4.8 \pm 0.2$ \\
$a_1(1260)$ &  $4.3 \pm 0.2$ \\
$b_1(1235)$ &  $3.8 \pm 0.2$ \\
$a_0(980)$ &  $3.0 \pm 0.1$ \\ \hline
Sum: &  $83.7 \pm 0.7$ 
\end{tabular}
\end{table}

%\begin{figure}[b]
%\vspace*{\fill}
%\centerline{\psfig{figure=/homeb/spieles/charm/work/collspec/result/ratesaqm.eps,width=12cm}}
%\caption{
%$J/\psi$ dissociation rates due to collisions with $\pi$'s and $\rho$'s
%in central Pb(160~GeV)+Pb reactions for different 
%choices of dissociation cross sections.
%The thick lines indicate the rates for the fixed cross section (see text).
%The thin lines show the result when the (energy dependent)
%cross sections from \protect\cite{mueller} are employed.
%\label{ratesaqm}}
%\vspace*{\fill}
%\end{figure}

\end{document}